\documentclass[doublecol]{epl2} 

\title{Entanglement entropy of aperiodic quantum spin chains}
\shorttitle{Entanglement entropy of aperiodic quantum spin chains}
\usepackage{graphicx}
\usepackage{epsfig}
\usepackage{multirow}
\bibstyle{apsrev.bib}
\author{Ferenc Igl\'oi\inst{1,2} \and R\'obert Juh\'asz\inst{1} 
\and Zolt\'an Zimbor\'as\inst{3}}
\shortauthor{F. Igl\'oi \etal}

\institute{
 \inst{1} Research Institute for Solid
State Physics and Optics, H-1525 Budapest, P.O.Box 49, Hungary  \\              \inst{2} Institute of Theoretical Physics,
Szeged University, H-6720 Szeged, Hungary \\
\inst{3} Research Institute for Particle and Nuclear Physics, 
H-1525 Budapest, P.O.Box 49, Hungary
}
\pacs{75.10.Jm}{Quantized spin models}
\pacs{03.65.Ud}{Entanglement and quantum nonlocality}
\pacs{03.67.Mn}{Entanglement production, characterization, and manipulation}
\pacs{75.50.Kj}{Amorphous and quasicrystalline magnetic materials}

\abstract{
We study the entanglement entropy of blocks of contiguous spins in
non-periodic (quasi-periodic or more generally aperiodic) critical 
Heisenberg, XX and quantum Ising spin chains, e.g. in Fibonacci chains.
For marginal and relevant aperiodic modulations, the entanglement entropy is 
found to be a logarithmic function of the block size with log-periodic
oscillations. The effective central charge, $c_{\rm eff}$, defined
through the constant in front of the logarithm may depend
on the ratio of couplings and can even exceed the corresponding value 
in the homogeneous system. In the strong modulation limit, the ground state 
is constructed by a renormalization group method
and the limiting value of $c_{\rm eff}$ is exactly calculated.
Keeping the ratio of the block size and the system size constant, the
entanglement entropy exhibits a scaling property, however, the corresponding scaling function may be nonanalytic.}

\begin{document}

\maketitle

\newcommand{\bc}{\begin{center}}
\newcommand{\ec}{\end{center}}
\newcommand{\be}{\begin{equation}}
\newcommand{\ee}{\end{equation}}
\newcommand{\beqn}{\begin{eqnarray}}
\newcommand{\eeqn}{\end{eqnarray}}

\section{Introduction}

Recently, the entanglement properties of strongly correlated systems 
have attracted great attention both in condensed 
matter physics\cite{calabrese_cardy} and in quantum information 
theory\cite{vidal}, for a review see\cite{fazio}. 
Much work is devoted to homogeneous one-dimensional systems in 
which the entanglement entropy of $L$ contiguous spins,
defined as the von Neumann entropy of the density matrix of the block,
$S_L= -{\rm Tr} \rho_L \log_2 \rho_L$, scales at the 
critical point as:
\be
S_L= \frac{c}{3} \log_2 L + k,
\label{S_L}
\ee
where $c$ is the central charge of the associated conformal field theory
and $k$ is a non-universal constant. In the vicinity of the quantum critical 
point, $L$
in Eq.(\ref{S_L}) is replaced by the correlation length \cite{calabrese_cardy}.

Inhomogeneities of different kinds (localized or extended defects, quenched
disorder, etc.) are able to modify the local critical behavior of the 
system\cite{ipt} and as a consequence,
the scaling properties of $S_L$ in Eq.(\ref{S_L}) can also be changed.
We expect that in case of
weak perturbations caused by inhomogeneities
the entropy scaling remains invariant or is modified
depending on the stability of the local critical
behavior at the boundary of the block. 
For {\it irrelevant perturbations} the critical properties
of the system at the boundary of the block remain unchanged and
it is natural to assume that  
the same scaling law holds for the entanglement entropy as in 
the homogeneous system.
On the contrary, for {\it relevant perturbations}
the local critical behavior at the block boundary is governed by a 
new fixed point and consequently the scaling
of the entanglement entropy is expected to be modified.
Finally, for {\it marginal perturbations}
the local critical behavior is characterized by continuously 
varying local exponents, thus, 
also the prefactor of the logarithmic entropy scaling
is presumably a continuously varying function of the strength of the inhomogeneity.

This scenario has been checked\cite{XXZ_defect} for a single defect coupling
located at the boundary of the block in the spin-$\frac{1}{2}$ XXZ chain 
defined by the Hamiltonian
\be 
H_{XXZ}=\sum_{i=1}^N J_i (S_i^xS_{i+1}^x+S_i^yS_{i+1}^y+\Delta S_i^zS_{i+1}^z),
\ee
where the $S^{\alpha}_i$'s ($\alpha=x,y,z$) are spin-$\frac{1}{2}$ operators and
$J_i=J$ for $i \ne L$ and $J_L \ne J$.
Concerning the behavior of the entanglement entropy, 
this type of perturbation is found to be irrelevant in 
the ferromagnetic domain ($\Delta<0$)
and relevant in the antiferromagnetic (AF) domain 
($\Delta>0$) in complete agreement with
the local critical behavior of the system. 
More interestingly, 
the boundary defect is a (truly) marginal perturbation in the $XX$
chain (i.e. with $\Delta=0$), where the logarithmic scaling form in
Eq. (\ref{S_L}) is still valid, although $c$ is replaced by  
a so called effective central charge,
$c_{\rm eff}$, which is found to depend continuously on
the strength of the defect \cite{XX_defect}. 
We have observed\cite{qic} a similar marginal behavior
with a continuously varying effective central charge in case of a single defect in the critical quantum Ising chain (QIC) defined
by the Hamiltonian
\be
H_I=-2 \sum_{i=1}^N J_i S_i^x S_{i+1}^x - h \sum_{i=1}^N S_i^z
\label{H_I}
\ee
with $J_i=J$ for $i \ne L$ and $J_L \ne J$ \cite{critbev}.

Quenched random disorder is a relevant perturbation both for the QIC 
and for the AF $XXZ$ model\cite{harris}. 
An arbitrarily small random perturbation drives the system
to an infinite randomness fixed point (IRFP), which can be studied in the
framework of an asymptotically exact renormalization group (RG) method
\cite{mdh,fisherxx,review}. In accordance with this, it was
found that the effective central charge
jumps to the value characteristic for the IRFP for
an arbitrarily weak random modulation \cite{refael,nicolas,calabrese}.

The effect of non-periodic (quasi-periodic or more generally
aperiodic) disorder has many similarities to the effect of
quenched random perturbations. After the discovery of 
quasi-crystals\cite{q_crystal}, aperiodic systems have become the subject
of intensive studies, both experimentally and
theoretically\cite{q_review,luck_rev}. 
Contrary to random disorder, these types of perturbations may also be
irrelevant or marginal.
The relevance of an aperiodic perturbation on the critical behavior 
of a quantum chain
is related to the sign of the cross-over exponent\cite{harris-luck}, 
$\phi=1+\nu(\omega-1)$, where $\nu$ is the
correlation length critical exponent of the pure chain and 
$\omega$ is the wandering exponent
of the aperiodic model \cite{wandering}.
According to a heuristic criterion \cite{harris-luck} 
the perturbation is relevant if $\phi>0$, marginal if $\phi=0$
and irrelevant if $\phi<0$. 
Later aperiodic QIC-s\cite{it1,it2,hgb}
and XX chains\cite{hermisson} were studied by an RG method and the 
low-energy properties were
exactly calculated. 
For aperiodic XXX chains, 
field-theoretical methods\cite{ft} and a variant of the
strong disorder RG approach have been 
applied\cite{hida1,hida2,vieira1,vieira2}. 
For all models with marginal or relevant
perturbations anisotropic scaling behavior is observed
which manifests itself in the scaling of the energy gap $\epsilon \sim N^{-z}$,
where the dynamical exponent $z$ is greater than one.

In this paper, we study the entanglement properties of critical 
aperiodic quantum spin chains
as a function of the strength of aperiodicity, measured by the ratio of 
the different couplings, $r$.
We investigate systematically the effect of different type of
non-periodic perturbations, such as irrelevant, relevant or marginal 
ones and examine the scaling form of the entropy in these critical, but
not conformally invariant systems. It is known that the spectrum of aperiodic 
systems have many unusual features\cite{q_review}
and we are interested in how these singularities are reflected in 
the entanglement properties. Spin chains both with continuous 
symmetry (XX and XXX models) and with discrete
symmetry (QIC) are considered. 

\section{Methods}
\begin{table}[h]
 \begin{tabular}{|c|c|cc|cc|cc|}  \hline
seq. & $r$ & \multicolumn{2}{|c|}{QIC}  & \multicolumn{2}{|c|}{XX} & \multicolumn{2}{|c|}{XXX} \\ \hline
  & $1$ & {\bf I} &  & {\bf M} & & {\bf R} & \\
Fib.  &  &  & $FF$ & & $FF$  &  & \\
  & $0$ & {\bf I} &  & {\bf M} & $RG$ & {\bf R} & $RG$\\ \hline
  & $1$ & {\bf M} &  & {\bf M} & & {\bf R} & \\
Tripl.  &  &  & $FF$ & & $FF$  &  & \\
  & $0$ & {\bf M} &  & {\bf M} & $RG$ & {\bf M} & $RG$\\ \hline
  & $1$ & {\bf M} &  & {\bf D} & & {\bf D} & \\
P.D.  &  &  & $FF$ &  &        &  & \\
  & $0$ & {\bf M} & $RG$ & {\bf D} &  &{\bf D}  & \\ \hline
  & $1$ & {\bf M} &  & {\bf D} &  & {\bf D} & \\
Hier.  &  &  & $FF$ &  &  &   & \\
  & $0$ & {\bf M} & $RG$ &  {\bf D} &  & {\bf D} &  \\ \hline
\end{tabular}
\caption{Summary of sequences (Fib. - Fibonacci, Tripl. - tripling,
P.D. - period doubling, Hier. - hierarchical) studied for different models (QIC, XX and XXX)
in this paper. The relevance of the aperiodic perturbation at the fixed points
($r=1$ pure system, $r=0$ extreme aperiodic system) 
are indicated as {\bf I} - irrelevant, {\bf M} - marginal,
{\bf R} - relevant, whereas letter 
{\bf D} refers to a noncritical dimerized ground state. 
The applied methods of investigations in the given range of $r$
are also shown as: $FF$ - free-fermion numerical calculation, 
$RG$ - analytical renormalization group study. \label{table:1}}
\end{table}
In the case of QIC and XX models, which can be mapped to a system
of free-fermions by means of standard techniques \cite{lsm}, 
we have performed large scale numerical
calculations.
Here, one determines first the restricted correlation matrix,
$G_{i,j}^L$, $i,j=1,2,\dots L$, which is the corresponding $L \times L$
minor of the matrix defined in Eq.(2.32c) in Ref.\cite{lsm}.
Then one determines the eigenvalues $\nu_l^2$ ($l=1,2,\dots L$) of the
positive definite symmetric matrix\cite{qic} $G^L (G^L)^T$, 
where $(G^L)^T$ denotes the transpose of $G^L$. 
In this representation, the entanglement entropy is given as a sum of
binary entropies of the non-correlated fermionic modes\cite{peschel,vidal}:
\be
S_L(N)=\sum_l - \lambda_l \log_2(\lambda_l)-(1-\lambda_l) \log_2(1-\lambda_l),
\label{S_binary}
\ee
where $\lambda_l=(1+\nu_l)/2$.

In the actual calculations, we considered blocks of size $L$ which
are given by finite approximants of the aperiodic sequence in order to get rid
of log-periodic oscillations. The complete system then consists of $N=2L$
spins with periodic boundary conditions and the entropy is generally
averaged over the $L$ different starting positions of the block.

For the aperiodic sequences we used in this paper, we also consider the strong aperiodic
modulation limit, $r \to 0$ or $1/r \to 0$, when
the average entanglement entropy is studied by a variant of the strong
disorder RG method. For the aperiodic AF XX and Heisenberg chains, the ground state in this
limit is of an
aperiodic singlet form, which is analogous to the random singlet phase of disordered
chains. The entanglement entropy is then given by the number of singlet bonds connecting the
block with the rest of the system. For the strongly aperiodic QIC, the ground state is of an
aperiodic embedded cluster form and the entropy is obtained by
counting the clusters connecting the block with the rest of the system. The implementation of the RG method
is described in details where the specific problems are treated.
The aperiodic sequences we study in this paper are selected in such a way to illustrate
the general properties of the entropy in aperiodic quantum chains. 
The considered sequences are summarized in
Table~\ref{table:1}, in which we have also given the stability of the
two fixed points: the pure
system's fixed point ($r=1$) and the strongly aperiodic system's
fixed point ($r=0$).
The applied methods used for the different models in the given range of the ratio $r$ are
also indicated.

\section{Fibonacci modulation}
\begin{figure}[h]
\onefigure[width=1. \linewidth]{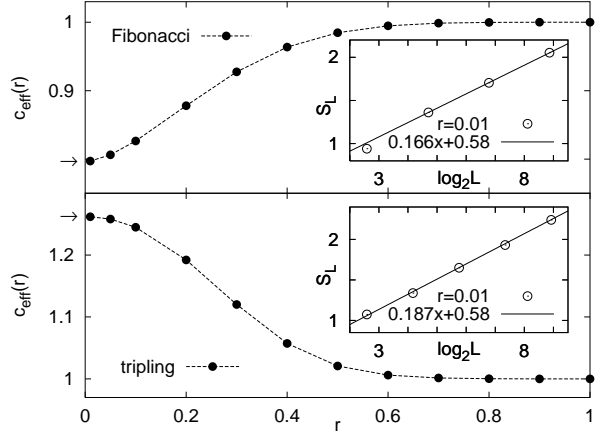}
\caption{Upper panel: average effective central charge of the Fibonacci XX chain as
a function of the coupling ratio, $r$. In the inset $S_L$ vs. $\log_2 L$ is shown for the QIC at $r=0.01$.
Lower panel: the same for the tripling sequence.}
\label{marg}
\end{figure}

Many basic features of entropy scaling can be seen for the Fibonacci
modulation, which
is irrelevant for the QIC ($\nu=1, \omega=-1$), marginal for the 
XX chain ($\nu=1, \omega=0$) and relevant for the XXX chain ($\nu=2/3,
\omega=0$).
The Fibonacci sequence, which consists of two different letters $a$ and $b$, 
is defined by the inflation rule: $a \to ab$ and $b \to a$, so that we have 
by iteration: $a,~ab,~aba,~abaab,\dots$, and the length of the sequence in the
$l$-th iteration is the Fibonacci number, $F_l$. In the quantum chains, 
the couplings take
two values, $J_a=r$ and $J_b=1$, depending on the underlying letter.
To check the effect of an irrelevant perturbation for the QIC, we have calculated the size-dependence of the
entropy for a strong modulation, $r=0.01$, and plotted the results in the inset of the upper panel of
Fig.\ref{marg}. We can see that the entropy has the same scaling form
as in the homogeneous system
with a central charge, $c=0.500(1)$. Hence this value is found indeed
to be independent of the value of $r$.

Repeating the same type of calculation for the Fibonacci XX model, the effective central charge is found to be
$r$ dependent, see the extrapolated values in the upper panel of Fig.\ref{marg}, which is in agreement
with the marginal nature of the perturbation for this system. $c_{\rm eff}(r)$
is monotonously decreasing with decreasing $r<1$ and approaches a finite
limiting value at $r=0$ with a correction of $O(r^2)$.

We have calculated $\lim_{r\to 0}c_{\rm eff}(r)\equiv c_{\rm eff}(0)$
for the XX Fibonacci chain exactly
by a variant of the strong disorder RG approach\cite{review}. For details of the application
of the method, we refer to Refs.\cite{hida1,hida2,vieira1,vieira2}.
The basic idea is that two
spins coupled by a very strong bond, $J_b$, form a maximally entangled singlet and the words $aba$ and
$ababa$ are renormalized into the letters $b'$ and $a'$, representing 
the new effective couplings $J'_b \approx \kappa r^2$ and
$J'_a \approx \kappa^2 r^3$, respectively. For the XX model, the prefactor is $\kappa=1$. We note that this renormalization step
corresponds to the reversed triple application of the 
inflation transformation described earlier.
The renormalized infinite chain is again
a Fibonacci chain with $J'_a/J'_b=r$, so that one can repeat the transformation and finally
the ground state of the system consists of a set of singlet pairs. The entanglement entropy
in this case is given by the number of singlet bonds connecting the block with the rest of the
system\cite{refael}.
The effective central charge can be determined by calculating the
difference  $\Delta S=S_L-S_{L'}$ between the averages of 
entanglement entropies belonging to blocks of length $L=F_{l+3}$ and 
blocks of length $L'=F_{l}$.
In order to do this, we notice that 
$L/L'=\rho = \tau^3$ for large $l$, where
$\tau=(1+\sqrt{5})/2$ is the golden-mean ratio. The ratio of the length of
the renormalized letters is $\lambda(a')/\lambda(b') = \tau$ and the
ratio of the density of
the letters is given by $\mu(a')/\mu(b')= \tau$, thus the singlet
bonds represented by letter $b'$,
cover a fraction $n_{cov}=1/(\tau^2+1)$ of the chain. In this way, we obtain
\be
\Delta S= \frac{c_{\rm eff}(0)}{3} \log_2 \rho=2 n_{cov},
\label{S_Fibb}
\ee
and $c_{\rm eff}(0)=2/[(\tau^2+1)\log_2 \tau]=0.7962$ 
which is in excellent agreement 
with the numerical findings (see the
arrow in the upper panel of Fig.\ref{marg}).
A more detailed derivation of $c_{\rm eff}$ based on the
distribution of singlet lengths in the infinite system can be found in
Ref. \cite{aper_singlet}.  

For the XXX-chain the Fibonacci modulation is a relevant perturbation
and the fixed point at
$r=0$ is strongly attractive. Indeed, 
the RG-procedure described for the XX-model leads
to the same form of renormalized couplings but with a prefactor
$\kappa=1/2$, thus the ratio $J'_a/J'_b= r/2<J_a/J_b$ tends to zero 
even for a finite starting
value of $r$. Consequently, the effective central charge 
is that given in Eq.(\ref{S_Fibb}) and it is independent
of the coupling ratio $r$ (provided $r \ne 1$).

\section{Tripling modulation}
The next sequence we consider is a tripling sequence, which is 
(for small perturbations)
marginal both for the QIC and
for the XX model ($\nu=1, \omega=0$) and relevant for the XXX model
($\nu=2/3, \omega=0$).
This sequence consists of three different
letters, $a$, $b$ and $c$ and is defined by the inflation rule: $a \to aba$, $b \to cbc$ and
$c \to abc$. In the following, we consider $J_b>J_c>J_a$ and use the
parameterization: $J_c/J_b=J_a/J_c=r$.
For the QIC, the size-dependence of the entropy for $r=0.01$ is shown
in the inset of the
lower panel of Fig.\ref{marg}, which is characterized by a 
central charge $c_{\rm eff}=0.560(5)$.
For the XX model, the shift in the effective central charge is even
greater, see the extrapolated values in the lower panel 
of Fig.\ref{marg}. Surprisingly, this type of inhomogeneity enhances 
the entanglement as opposed to random and single defect perturbations 
in these models. 
We mention that in some special chains of spins with many components
also random disorder can enhance entanglement \cite{santachiara}.
In the XX case, we have also calculated the small 
$r$ limiting behavior by the 
RG method, in which singlets form over the $J_b$ bonds and the
new letters are defined by the deflation rules. 
The new couplings satisfy the marginality
condition: $J'_c/J'_b=J'_a/J'_c=r$ and in the aperiodic singlet ground state
we have $\rho=3$ and $n_{cov}=1/3$, such that we obtain from Eq.(\ref{S_Fibb}):
$c_{\rm eff}(0)=2/\log_2 3=1.262$ (see the arrow in the lower panel of 
Fig.\ref{marg}).
For the XXX chain, the tripling perturbation plays a special role: a weak perturbation at $r=1$
is relevant but it is marginal at $r=0$,
where the same type of condition: $J'_b/J'_c=J'_c/J'_a=r$ holds as for the XX chain.
Consequently, the effective central charge of the tripling XXX chain is expected
to be a continuously varying function of $r$ at least for small $r$ but discontinuous at $r=1$.
We note that the Fibonacci and the tripling XX (and XXX) chains belong to a class 
of aperiodic chains for which the ground state is an aperiodic singlet
state in the limit $r \to 0$ and
the average entropy can be exactly calculated\cite{aper_singlet}.

\section{Period doubling modulation}

In the rest of the paper, we focus on marginally non-periodic QIC-s for 
which even the first correction
to the entanglement entropy at $r=0$ is analytically calculated, thus we can 
gain more insight into the varying
nature of the effective central charge. 
First, we consider the 
period-doubling sequence, which
has the inflation rule $a \to ab$ and $b \to aa$, such that we have by iteration: $a$, $ab$, $abaa$, $abaaabab$, $abaaabababaaabaa$, $\dots$.
A small perturbation of this type 
is marginal for the QIC ($\nu=1,\omega=0$), whereas in case of the XX
and XXX chains, it drives the system into a non-critical phase with a
dimerized ground state \cite{hermisson}.   
For the critical QIC, we use the
parameterization: $h_c=1$, $J_a=r^{-1/3}$ and $J_b=r^{2/3}$, when the dynamical 
exponent is given $z=\ln(r^{2/3}+r^{-2/3})/\ln 2$ \cite{it1,it2}. 
The numerically calculated effective
central charge is shown in the inset of Fig.\ref{hier}, which is found to 
be $r$-dependent.
We have checked numerically that $c_{\rm eff}(r)=c_{\rm eff}(1/r)$.
The limiting value is $c_{\rm eff}(0)=0$, which is approached in a singular way
as opposed to the smooth behavior of a single defect problem\cite{XX_defect}.
 
The limiting behavior for $1/r \to 0$ is studied in the frame of 
an RG approach\cite{szallas},
in which one transformation step consists of two parts. 
In the first part, all strong $J_b$ bonds
are eliminated and two-site clusters with an effective field,
$\tilde{h} \approx h_c^2/J_b=r^{-2/3}$, are created.
In the second part, all the strong transverse fields with $h_c=1$ are eliminated and new couplings $\tilde{J_a} \approx J_a^3/h_c^2 = r^{-1}$ are created. Identifying $\tilde{J_b}$ with
$J_a=r^{-1/3}$ we have an aperiodic sequence
$\tilde{a} \tilde{b}\tilde{b}\tilde{a}\tilde{a}\tilde{a}\tilde{b}\tilde{b}\tilde{a}\tilde{b}\tilde{b}\dots$
which is in a period doubling form in terms of $a'\equiv\tilde{a}$ and $b'\equiv\tilde{b}\tilde{b}$.
During further renormalization steps the structure of this chain remains invariant. To calculate
the entanglement entropy, we consider a chain with $N=4^n$ sites, in
which there is a strong bond at $L=N/2$.
By eliminating this bond, an effective cluster is formed, which connects the block with the rest of the
system, such that in the limit $1/r\to 0$ $S_L(N)=1$ and $c_{\rm eff}(0)=0$ in accordance with the
numerical results presented in Fig.\ref{hier}. In the free-fermion representation, this result
follows from Eq.(\ref{S_binary}) in which there is only one mixed non-correlated fermionic mode
in the subsystem with a non-zero eigenvalue $\lambda_1=1/2$ (i.e. $\lambda_l=0$, for $l=2,3,\dots L$).
The perturbative correction to this term
for $1/r \ll 1$ can be estimated as follows. 
For $n=2$, i.e. for $N=16$, we obtain the sequence $\tilde{a} \tilde{b}\tilde{b}\tilde{a}$ after the first RG step, while by
eliminating the two strong $\tilde{b}$ bonds in the second RG step,
a super-cluster is formed over the boundary of the block. 
The entanglement contribution due to this super-cluster is calculated perturbatively leading to a new
mixed fermionic mode with non-zero eigenvalue:
$\lambda_2=r^{-2/3}/16$. For $n=3$, i.e. for $N=64$, three RG steps
can be performed, which results in
a new super-super-cluster and a new mixed mode, so that there are then two non-zero
sub-leading eigenvalues: $\lambda_2=\lambda_3=r^{-2/3}/16$. 
For a general $n$, we then have $n$ embedded clusters and there are
$n-1$ non-zero sub-leading eigenvalues:
$\lambda_2=\lambda_3= \dots=\lambda_n=r^{-2/3}/16$. The entanglement entropy for small $1/r$ indeed
scales with $n-1=\log_2 N/2-1$ and the effective central charge is given by:
$c_{\rm eff}(r)=r^{-2/3}(\log_2 r/8+{\rm const})$.
The symmetry of the effective central charge $c_{\rm eff}(r)=c_{\rm
  eff}(1/r)$ observed numerically can also be seen in the perturbative
treatment for small $r$. Hence, the function $c_{\rm eff}(r)$ is
indeed singular at $r=0$ in accordance with the numerical results in the inset of Fig.\ref{hier}.

\section{Hierarchical modulation}

The last sequence we consider in this paper is the hierarchical one in
which the couplings are given by\cite{huberman}:
\be
J_i=Jr^n,\quad i=2^n(2m+1),\quad n,m=0,1,\dots
\ee
The hierarchical sequence is limit periodic\cite{kls} and can be generated through substitution
with an infinite alphabet. For the QIC, the critical point is located at \cite{critbev}
$h_c=Jr$ and by setting $J=1/r$, we have $h_c=1$.
The excitation energy scales as $\epsilon \sim N^{-z}$, with a
dynamical exponent $z=\ln(r+1/r)/\ln 2$ \cite{it1,it2}. (For the XX and
XXX models this type of modulation drives the system to a dimerized phase.)

\begin{figure}[h]
\onefigure[width=1. \linewidth]{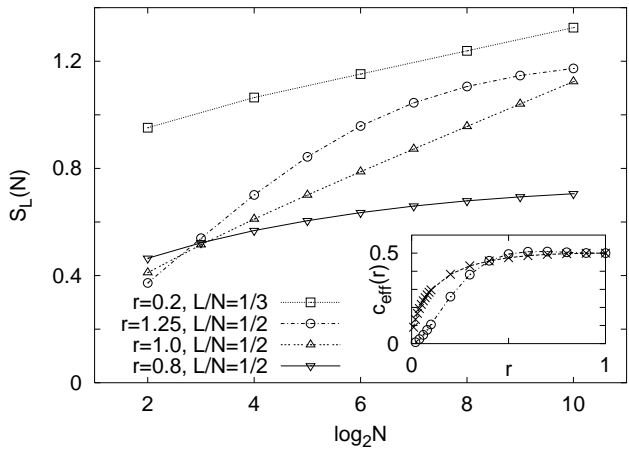}
\caption{Size dependence of the entanglement entropy of the
  hierarchical QIC for three different values $r$ at $L/N=1/2$ and for $r=0.2$ at $L/N=1/3$.
Inset: extrapolated effective central charges as a function of $r$. 
Hierarchical chain at $L/N=1/3$ ($\bigcirc$), period-doubling chain at
$L/N=1/2$ ($\times$).}
\label{hier} 
\end{figure}

In the numerical study we considered open Ising chains of 
length $N=2^n$ and calculated
the entanglement entropy of the half of the system, i.e. with
$L=N/2$. As can be seen in Fig.\ref{hier}, the entropy approaches a
finite limiting value for any $r \ne 1$. Saturation of the entropy is found also
at $L=N/4$ and $L=N/8$ and expected to hold for any $L=N/2^{p}$, $p=1,2,\dots$,
since at these special positions there are very small couplings, which act in the
infinite system as an effective cut. Repeating the calculation with a block of size
$L=[N/3]_{int}$, where $[y]_{int}$ denotes the integer part of $y$, we obtain the
usual logarithmic dependence as can be seen in Fig.\ref{hier}. In this case $c_{\rm eff}(r)$
is found to be a continuously varying function of $r$ (see the extrapolated values in the inset of Fig.\ref{hier}).
We have thus the surprising conclusion that the scaling function $\tilde{S}(x=L/N)=\lim_{N \to \infty} S_L(N)/\log_2 N$ is non-analytic at
$x=2^{-p}$, $p=1,2,\dots$, in contrast
to the behavior in homogeneous systems\cite{calabrese_cardy}.

Next, we turn to a perturbative RG treatment for small $r$ 
and focus on the block composed of the first $L$ spins at the left
boundary but now we consider the thermodynamic limit $N \to \infty$. 
As for the period doubling modulation, we
eliminate the strongest bonds, which 
are at odd positions (i.e. with $n=0$), and create renormalized
two-spin clusters subjected to an effective transverse field
of $\tilde{h} \approx h_c^2/J=r$. The renormalized
chain has now $N/2$ sites with the same hierarchical structure of couplings.
In the repeated use of the transformation, bonds with $n=1$ (then with $n=2$, etc)
are decimated out and super-clusters with a hierarchical structure are created. Then for any value of $L$, the
block is connected to the environment through one super-cluster, thus the entanglement entropy is $S_L=1$
and $c_{\rm eff}(0)=0$ in accordance with the results presented in the inset of Fig.\ref{hier}. This result can also be derived using the
free-fermion representation of the system, when in the limit $r \to 0$,
the density matrices of the non-correlated fermionic modes have
eigenvalues ($\lambda_i$, $1-\lambda_i$) given by $\lambda_1=1/2$ and $\lambda_l=0$, for $l=2,3,\dots L$, thus $S_L=1$.

For a small but finite $r$, there are perturbative corrections, which depend on the value of $L$.
In the simplest case $L=3$, these corrections are due to
correlations between the spin-clusters $(1,2)$ and $(3,4)$. In the
non-correlated free-fermionic description, the eigenvalues are up to $O(r^2)$: $\lambda_1=1/2+O(r^2)$, $\lambda_2=r^2/4$
and $\lambda_l=0$, for $l=3,\dots L$.
For $L=5$ there are two perturbative cluster-cluster contributions which will result in $\lambda_2=2r^2/4$,
whereas the other eigenvalues remain unchanged, at least up to $O(r^2)$. Generally, for $L_{\kappa}=2^{\kappa}+1$,
$\kappa=0,1,\dots$ there are $\kappa$ perturbative contributions and in the limit
$\kappa r^2 \ll 1$ we have $\lambda_2=\kappa r^2/4$. (cf. the form of
perturbative cluster correction in the case of the period doubling sequence.)
Thus we obtain for the entanglement entropy:
\be
S_{2^{\kappa}+1}=1+\kappa r^2 \left[ -\frac{1}{4} \log_2(\kappa r^2)+\frac{1}{2}+\frac{1}{4\ln 2}
\right]+O(\kappa^2 r^4)
\label{kappa}
\ee
For a general value of $L=2^n(2m+1)$, the number of independent perturbative
(cluster) contributions, $\kappa$, does not depend on $n$ and given by $\kappa=[\log_2 m]_{int}+1$.
In this case the entanglement entropy
in the leading order is given in Eq.(\ref{kappa}).
We thus conclude, that $S_L(r)$ has log-periodic oscillations in
$L$ and the leading
$L$ dependence is faster than logarithmic in the range $(\log_2
L) r^2 \ll 1$, although for $(\log_2 L) r^2 \gg 1$, a logarithmic
 $L$ dependence is expected.

\section{Summary}
Summing up, we have studied the entanglement properties
of critical quantum spin chains, such as the antiferromagnetic
Heisenberg and XX models,
as well as the quantum Ising chain 
in the presence of aperiodic modulations of the
couplings. As far as the critical properties of the
systems are concerned, an aperiodic inhomogeneity can be irrelevant, 
marginal or relevant, depending on the model and on the fluctuation 
properties of the aperiodic sequence. An irrelevant aperiodicity
alters neither the logarithmic scaling of the entropy nor 
the constant in front of the logarithm, i.e. the growth of the entropy
is determined by the central charge $c$ of the pure system.
For marginal and relevant modulations, the average entropy is
still scaling logarithmically with the size of the block, however,
the prefactor is modified compared to that of the homogeneous system
i.e. $c_{\rm eff} \ne c$ and in addition to this, log-periodic
oscillations appear.
Interestingly, $c_{\rm eff}$ can exceed the corresponding value in 
the homogeneous system
and for marginal perturbations, it is a continuous 
function of the coupling ratio $r$.
In the $r \to 0$ limit, where the ground state of several aperiodic
sequences can be exactly constructed, the average effective central charge 
was analytically calculated.

\acknowledgments
This work has been supported by the National Office of Research and
Technology 
under
Grant No. ASEP1111, by a German-Hungarian exchange program (DAAD-M\"OB), by the
Hungarian National Research Fund under grant No OTKA TO43159,
TO48721, K62588, MO45596 and M36803.
RJ acknowledges support by the Deutsche Forschungsgemeinschaft
under Grant No. SA864/2-2.




\begin{thebibliography}{0}
\bibitem{critbev}
The critical external magnetic field $h_c$ of this model is given by
$\ln h_c=\overline{\ln J}$, where $\overline{(\dots)}$ stands for the 
average in the infinite system,
\Name{P. Pfeuty}
\REVIEW{Phys. Lett}{72A}{1979}{245}.
\bibitem{wandering}
For the QIC local energy fluctuations are given by $\Delta J_i=J_i-\overline{J}$,
whereas for the AF $XX$ and Heisenberg models the local energy fluctuation is the dimerization:
$\Delta J_i=J_{2i-1}-J_{2i}$. Accumulated energy fluctuations have the scaling
behavior: $\sum_{i=1}^{N} \Delta J_i \sim  N^{\omega}$. For $\omega<1$ the system stays critical,
and we study here only this type of sequences. Note that the same
aperiodic sequence could be characterized by different wandering exponents for the different models.
\end{thebibliography}
\end{document}